\documentstyle[aps,multicol,epsf]{revtex}

\begin{document}
\draft

\title{Stratification Instability in Granular Flows}
\author{Hern\'an A. Makse}
 
\address{
Laboratoire de Physique de la Mati\`ere Condens\'ee, Coll\`ege de
France\\ 
11 place Marcelin Berthelot, 75231 Paris Cedex 05, France\\ 
and \\
Center for Polymer Studies, and Physics Dept., Boston
University, Boston, MA 02215 USA
}
\date{Phys. Rev. E, {\bf 56}, 7008-7016 (1997)}
\maketitle
\begin{abstract}
When a mixture of two kinds of grains differing in size and shape is poured
in a vertical two-dimensional 
cell, the mixture spontaneously stratifies in alternating
layers of small and large grains, whenever the large  grains are more faceted
 than the small grains.
Otherwise, the mixture spontaneously segregates in different regions of
the cell when the large grains are more rounded than the small grains.
We address the question of the origin of 
the instability mechanism leading to stratification 
using a recently
proposed set of equations for 
surface flow of granular mixtures. 
We show that the stable solution of the system is a segregation
solution due to size (large grains  tend
to segregate downhill near the substrate
and small grains tend to segregate uphill) 
and  shape (rounded grains
tend to segregate downhill and more faceted grains tend to segregate uphill).
As a result, the segregation solution of the system is realized for
mixtures of large-rounded grains and small-cubic grains with the
large-rounded grains segregating near the bottom of the pile.
Stability analysis 
reveals the instability mechanism driving the system to stratification 
as a competition between size-segregation and shape-segregation taking place
for mixtures of large-cubic grains and small-rounded grains.
The large-cubic 
grains tend to size-segregate at the bottom of the pile, while at the 
same time, they tend to shape-segregate near  the pouring point.
Thus, the segregation solution becomes unstable, and the system
evolves spontaneously 
to  stratification.
\end{abstract}

\begin{multicols}{2}

\narrowtext
\section{Introduction}

One of the unusual properties \cite{bagnold,borges}
of granular materials 
\cite{review1,review2,review3,review4,review5,review6} 
is the size-segregation of
mixtures when they are
exposed to external periodic perturbations such as
vibrations or rotations  
\cite{brazilnut1,brazilnut2,brazilnut4,brazilnut4.5,brazilnut5,brazilnut8,brazilnut10}.
 Size-segregation 
also occurs  
when a mixture of grains of different
size
is simply poured onto a heap
\cite{segregation1,segregation2,segregation3,segregation4,segregation5,segregation6}:
the large grains spontaneously segregate near the bottom of the heap,
whereas	the small grains segregate near the pouring point at the top of
the heap.

Recently,
it was shown \cite{makse1}
that when a mixture 
composed of grains differing not only in size but also in shape
is poured  in a ``granular Hele-Shaw
cell'' (two
vertical slabs
separated by  a gap $\approx$ 5 mm), a spontaneous
stratification is observed.
Granular mixtures of
small-rounded grains and large-cubic
grains 
stratify in  alternating layers of small-rounded and large-cubic
grains parallel to the surface of the pile when they are poured in the cell.

According to the experiments \cite{makse1}, 
the control parameter for stratification appears to be the difference
of the repose angles of the pure species
\begin{equation}
\delta \equiv
\theta_{22} - \theta_{11},
\label{delta}
\end{equation}
where $\theta_{11}$ is the angle of repose of the small grains, and
$\theta_{22}$ is the angle of repose of the large grains.
The stratification experiments \cite{makse1} used a mixture of grains
of different shapes (small rounded or less faceted grains and large  cubic
or more faceted grains). The repose angle of the smaller pure species is then
smaller than the repose angle of the large pure species---i.e.,
$\delta > 0$. 
On the other hand, strong segregation but not stratification  
occurred \cite{makse1} when 
$\delta< 0$,  (corresponding to a
mixture of small-cubic  grains, and large-rounded
grains) \cite{ratio1}.
We 
notice that the angle of repose of the pure species does not depend on the
size of the grains, and it is a function of the shape of the grains: the
rougher the grains the larger the angle of repose.

To describe the case of a single-species sandpile in a two-dimensional
geometry, Bouchaud, Cates, R. Prakash, and Edwards (BCRE)  
\cite{bouchaud1,bouchaud2} developed a novel
theoretical approach.
They introduced 
two coarse-grained 
variables: the local height of the sandpile, and the local
``thickness''  of the layer of rolling grains, and a set of coupled
equations to govern the flow of the rolling grains
and their interaction with the sandpile. Recently, de Gennes
\cite{pgg} applied the BCRE formalism to the case of granular flows
in a thin rotating drum \cite{thin1,thin2},
and very recently
 Boutreux and de Gennes (BdG) \cite{bdg} treated
the case of granular flows made of two
species of different angles of repose.
Makse, Cizeau and Stanley (MCS) \cite{makse2,cms} 
reproduced stratification and segregation 
using a discrete model and the continuum approach developed by BdG.
They  showed that a  ``kink''
mechanism \cite{makse2,cms} describes  
the dynamics of stratification in agreement with experimental findings
\cite{makse1}.

In this paper, we address
the question of the origin of the instability leading to stratification.
We study analytically segregation and stratification 
as observed in 
\cite{makse1} when the two species have  different
size and different shape (or, in general, different angle of repose)
\cite{ratio2}. 
We use the 
continuum
approach of BdG \cite{bdg}, and 
MCS \cite{makse2,cms},
to calculate the steady-state solution of the equations
of motion for surface flow of granular mixtures. 
This solution shows the complete size-segregation of the mixture with
the large grains being found at the bottom of the pile.
We then study analytically the
conditions under which the instability leading to stratification occurs.
Stability analysis show 
that the  
steady-state solution 
is stable under
perturbations only when $\delta<0$, 
while the
segregation solution is unstable (leading to stratification)
when $\delta>0$, in agreement with experiments \cite{makse1}. 

\begin{figure}
\centerline{
\vbox{  \hbox{\epsfxsize=6.5cm \epsfbox{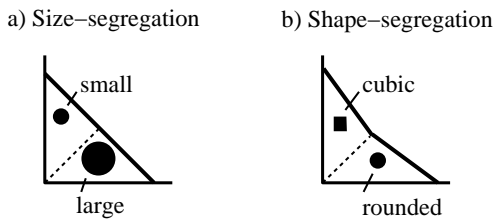}}
     }
       \vspace{1cm}    
}
\narrowtext
\caption{Two  segregation effects acting when the
 grains differ in size and
shape.
{\bf a,} Size-segregation:
large grains tend to segregate at  the bottom of the pile.
{\bf b,} Shape-segregation: rounded grains tend to
segregate at the bottom of the pile.} 
\label{origin}
\end{figure}

The stratification instability can be seen as follows.
There are two segregation mechanisms acting when pouring a mixture of grains
differing in size and shape in a cell:

\begin{itemize} 
\item
{\bf Size-segregation:} large grains 
tend to segregate downhill near the bottom of the pile, and  small grains tend
to segregate uphill near the pouring point, since large
grains roll down easier on top of small grains than small grains on top
of large grains (Fig. \ref{origin}a).

\item
{\bf Shape-segregation:} cubic grains tend to segregate near top of the
pile, and rounded grains segregate near the bottom, since rounded
grains
roll down easier than cubic grains (Fig. \ref{origin}b).
\end{itemize}

Thus,  when pouring a mixture  of small-cubic grains and
large-rounded grains the segregation of the mixture results, since the 
small-cubic grains size-segregate and shape-segregate near the top of
the pile, and the large-rounded grains size  and shape-segregate
near the bottom.
This situation gives rise to the steady-state solution of the system.

On the other hand, 
the stratification process arises as a consequence of an instability
mechanism.
For 
 mixtures of large-cubic  grains and  small-rounded grains
 there exists a competition between size-segregation and shape-segregation.
Large-cubic grains tend to size-segregate at the bottom
of the pile, while at the same time, they
tend to shape-segregate at the  top of the pile.
Thus, the segregation solution becomes unstable, and
the instability drives the system spontaneously to stratification.

In the following we take up each of these results in turn.
The paper is organized as follows: In Section \ref{theory} we present
the theoretical formalism 
for surface flows of granular mixtures. 
In Section  \ref{ssregime} we calculate the steady-state solution of the
problem. In Section \ref{stability} we perform a stability analysis, and
in Section \ref{instability} we discuss the instability
mechanism for stratification and we propose a
phase diagram for surface flows of granular mixtures
in light of these results.

\section{Theory for Surface Flow of Granular Mixtures}
\label{theory}

The theoretical study of surface flows of granular materials was
triggered by the works of BCRE \cite{bouchaud1,bouchaud2} and Mehta and
collaborators \cite{mehta}.
In a recent theoretical study for 
the case of a single-species sandpile
BCRE \cite{bouchaud1,bouchaud2}
proposed 
two coupled variables to describe
the  dynamics of two-dimensional sandpile surfaces:
the local angle of the sandpile $\theta(x,t)$ (or alternatively the
height of the sandpile $h(x,t)$) which describes the static phase
(i.e., the grains which belong to the pile),
 and the local thickness of the layer of
rolling grains $R(x,t)$ to describe the rolling phase (i.e., the grains
that are not part of the pile but roll downwards on top of the static
phase). 
BCRE also proposed a set of 
convective-diffusion
equation 
for the rolling grains,
which was later simplified by de Gennes \cite{pgg}:

\begin{equation}
\frac{\partial R(x,t) }{\partial t} = - v \frac{\partial R}{\partial x}
+ \Gamma(R,\theta),
\end{equation}
where $v$ is the downhill drift velocity of the grains along $x$, 
 assumed to be
constant in space and in time.
The interaction term $\Gamma$ 
takes into account the conversion of static grains into
rolling grains,
and vice versa. The simplest 
form of $\Gamma$ is \cite{bouchaud1,bouchaud2,pgg} 
 
\begin{equation}
\Gamma(R,\theta) = \gamma ~~ \left( \theta(x,t) - \theta_r \right) ~~R(x,t).
\end{equation}
Here  $\theta_r$ denotes the angle of repose (the 
maximum angle below which a rolling
grain is converted into a static grain 
\cite{bouchaud1,bouchaud2,bagnold2,jaeger}).
The rate $\gamma > 0$ has dimension of inverse time,
and $v / \gamma$ represents the length scale at which a
rolling grain 
will interact significantly with a surface at an angle
slightly above or below the angle of repose \cite{pgg}. 
For notational convenience we do not consider the difference between the
angle and the tangent of the angle, i.e. 
\begin{equation}
\theta(x,t) \equiv -\frac{\partial h}{\partial x}.
\label{tangent}
\end{equation}
The equation for $h(x,t)$ follows by conservation
\begin{equation}
\frac{\partial h(x,t)}{\partial t} = - \Gamma(R,\theta).
\end{equation}

\begin{figure}
\centerline { \vbox{ \hbox
{\epsfxsize=7.cm \epsfbox{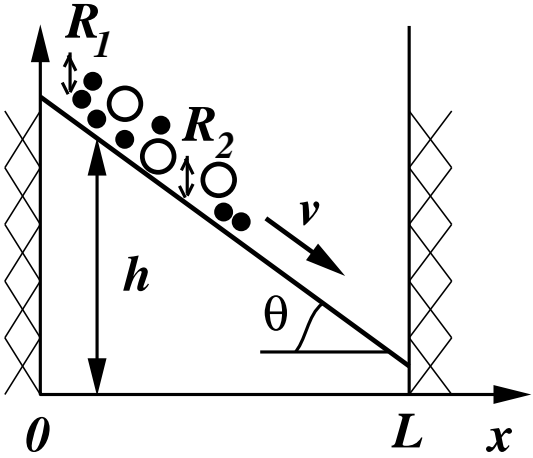} } }
}
\vspace{1cm}
\narrowtext
\caption{Diagram showing the variables used to describe
the granular flow of mixtures in a granular Hele-Shaw cell or 
two-dimensional silo.} 
 \label{coordinates}
\end{figure}

Recently, BdG
 \cite{bdg} have extended the BCRE formalism
to the case of two species.
This formalism considers the two local 
``equivalent thicknesses'' of the species in the rolling phase 
$R_{\alpha}(x, t)$ (i.e. the total  thickness of the rolling phase
 multiplied by the
local volume fraction of the $\alpha$ grains in the 
rolling phase at position $x$), with $\alpha =1,2$ respectively for small
and large grains.
The total thickness of the rolling phase is defined as 
\begin{equation}
R(x,t) \equiv
R_1(x,t) + R_2(x,t).
\end{equation}
The static phase is described by
 the height of the sandpile $h(x,t)$, and the volume
fraction of static grains $\phi_\alpha(x,t)$ of type $\alpha$ at the surface
of the pile.
Here $x$ is the longitudinal coordinate,
and the pouring point is assumed to be at $x=0$, and we consider a silo
or cell of lateral size $L$  (see Fig. \ref{coordinates}).

The equations of motion for the rolling species are \cite{bdg}
 
\begin{mathletters}
\label{bdg-eq}
\begin{equation}
\label{bdg-r}
\frac{\partial R_\alpha(x,t)}{\partial t}=-v_\alpha \frac{\partial
R_\alpha}{\partial x} +
\Gamma_\alpha,
\end{equation}
and the equation for $h(x,t)$ follows by conservation
 
\begin{equation}
\label{bdg-h}
 \frac{\partial h(x,t)}{\partial t}=- 
\Gamma_1-\Gamma_2.
\end{equation}

Here $v_\alpha$ is the downhill convection velocity of species
$\alpha$
along $x$.
The interaction term $\Gamma_\alpha$  
takes into account the conversion of rolling grains of type $\alpha$
into static grains, and the amplification of static grains $\alpha$ by
rolling
grains of type $\alpha$ or $\beta$. 
 $\Gamma_\alpha$ is  defined through a 
$2 \times 2$ collision matrix $\mbox{M}_{\alpha \beta}$
\begin{equation}
\Gamma_\alpha(\theta, R_\alpha, \phi_\alpha) 
\equiv \sum_{\beta=1}^{2} \mbox{M}_{\alpha
\beta} R_\beta. 
\end{equation}
The  elements of the collision matrix $\mbox{M}_{\alpha\beta}$ 
characterize the
interaction of a  rolling grain of type $\beta$ with a surface of static
grains of type $\alpha$, and they are determined
by the local angle 
$\theta(x,t)$, and
 the
concentrations
$\phi_\alpha(x,t)$.  
The concentrations  of static grains at the surface of the pile 
$\phi_\alpha(x,t)$ are given by
\begin{equation}
\phi_\alpha(x,t) \frac{\partial h}{\partial t} = - \Gamma_\alpha,
\label{bdg-phi}
\end{equation}
and
\begin{equation}
 \phi_1 + \phi_2 =1.
\end{equation}
\end{mathletters}

The  {\it canonical} form of the collision matrix is defined by 
taking into account a set of binary collisions between a rolling 
and a static grains \cite{bdg,exchange}

\begin{equation}
\label{canonical}
\hat{\mbox{M}} \equiv
\begin{array}{|lr|}
a_1(\theta) \phi_1 - b_1(\theta) & x_2(\theta) \phi_1 \\
x_1(\theta) \phi_2 & a_2(\theta) \phi_2 - b_2(\theta) 
\end{array}
\end{equation}
This definition involves 
a set of a priori unknown collision 
functions contributing to the rate processes:
$a_\alpha(\theta)$ is the contribution due to an amplification process,
(i.e., when a static grain of type $\alpha$ is converted into a rolling grain 
due to  
a collision by a rolling grain of type $\alpha$), $b_\alpha(\theta)$
is the contribution due to capture of a rolling grain of type $\alpha$,
(i.e., when a rolling grain of type $\alpha$ is converted into a static
grain), and
$x_\alpha(\theta)$ is the contribution due to a cross-amplification
process, (i.e., the amplification of a static grain of type $\beta$ due to
a collision by a rolling grain of type $\alpha$).

BdG \cite{bdg} used a ``minimal'' form of the 
collision matrix to  calculate
the steady-state solution 
in the geometry of a two-dimensional silo 
for the case of
mixtures of grains differing only in angle of repose. 
This solution shows a complete segregation at the
low edge of the silo, and near this point the concentrations of static
grains show power-law behavior. 
In their model, they consider a constant cross-amplification term
$x_\alpha(\theta)=$const, and they 
consider the angle of repose  of each species 
to be independent on the surface composition of the pile.
A generalization of the 
minimal model of \cite{bdg}
 to include not only different surface properties of the species
but also 
different size of the species with small size ratios $(d_2/d_1 < 1.4)$
is considered in \cite{bmdg}. 
 
Here,
we use the BdG equations to study segregation as well as
stratification of mixtures of grains differing in size and shape 
in a two-dimensional silo (Fig. \ref{coordinates}).
As in MCS \cite{makse2}, 
we focus on the dependence of the repose angle of every species
on the composition of the surface $\phi_\beta(x,t)$. We use the
following definitions of collision 
functions  \cite{makse2}:

\begin{mathletters}
\label{canonical2}
\begin{equation}
\begin{array}{llcl}
a_\alpha(\theta)&\equiv& \gamma_{\alpha\alpha} & \Pi[\theta(x,t)-
\theta_\alpha(\phi_\beta)] \\
b_\alpha(\theta)&\equiv& \gamma_{\alpha\alpha} &
\Pi[\theta_\alpha(\phi_\beta)-\theta(x,t)] \\
x_\beta(\theta) &\equiv& 
\gamma_{\beta\alpha} & \Pi[\theta(x,t) -\theta_\beta(\phi_\beta)],
\end{array}
\end{equation}
where 
 
\begin{equation}
\Pi[x] \equiv \left \{
\begin{array}{cr}
0  &~~~~~~~\mbox{if $x < 0$} \\
x  &~~~~~~~\mbox{if $x \ge 0$}
\end{array}
\right . .
\end{equation}
\end{mathletters}

\begin{figure}
\centerline { \vbox{ \hbox
{\epsfxsize=7.cm \epsfbox{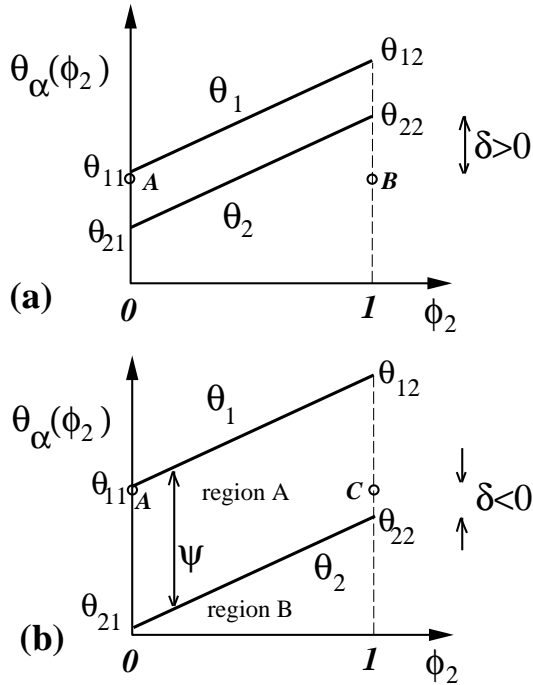} } }
}
\vspace{1cm}
\narrowtext
\caption{Dependence of the generalized  angle of repose
for the two types of rolling 
grains on the concentration of
the surface of large grains $\phi_2$ for {\bf a,} $\delta > 0$, and {\bf b,}
$\delta < 0$, where $\delta$ is defined in Eq. (\protect\ref{delta}). We
define $\theta_{\alpha \beta} = \theta_\alpha(\phi_\beta = 1)$
\protect\cite{makse2}.} 
\label{angles}
\end{figure}

Here, the rates are $\gamma_{\alpha \alpha} > 0$,
and the generalized  angle of repose
$\theta_\alpha(\phi_\beta)$ of a $\alpha$ type of rolling grain is a continuous
function of the composition of the surface
$\phi_\beta$ \cite{makse2} (see Fig.~\ref{angles}, where we 
also define 
$\theta_{\alpha\beta}$ as
$\theta_\alpha(\phi_\beta)$ for $\phi_\beta=1$):

\begin{equation}
\label{dependence}
\begin{array}{rcl}
\theta_1(\phi_2) &=& m \phi_2 + \theta_{11}\\
\theta_2(\phi_2) &=& m \phi_2 + \theta_{21} = - m \phi_1 + \theta_{22},
\end{array}
\end{equation}
where $m\equiv \theta_{12} - \theta_{11} = \theta_{22} - \theta_{21}$.
We have assumed the difference 
\begin{equation}
\psi \equiv
\theta_1(\phi_2)-\theta_2(\phi_2)
\end{equation}
 to be
independent of the concentration $\phi_2$.
We notice that 
the fact that grains 1 are smaller than
grains 2 implies $\theta_1(\phi_2) > \theta_2(\phi_2)$ for any given
concentration $\phi_2$ (i.e. the small grains are always the first to be
captured). The angular difference $\psi$ is thus determined by the
difference in size between the grains; 
the larger the size difference 
the larger $\psi$. Moreover, the fact that the grains
have different shapes implies $\theta_{11}\neq\theta_{22}$, (for
instance
$\theta_{11} < \theta_{22}$ when the grains $1$ are more rounded than
the grains $2$).

\begin{figure}
\centerline{
\vbox{ \hbox{\epsfxsize=7.cm
\epsfbox{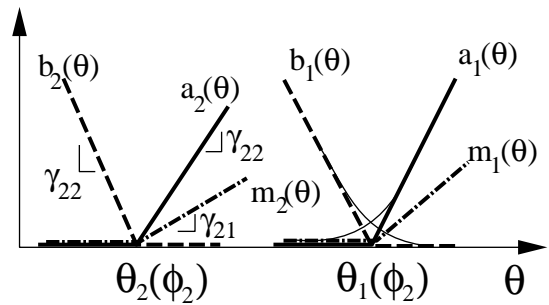}}
}}
\vspace{1cm}
\narrowtext
\caption{Schematic plot of capture $b_\alpha(\theta)$, 
amplification  $a_\alpha(\theta)$, and cross-amplification  $x_\alpha(\theta)$
functions used in the definition of the present model.
These functions are expected to be continuous in a region near
the angle of repose (as shown by the thin curves). However, when 
the ratio between the size of the grains is not close to one, then 
$\psi = \theta_1(\phi_2)- \theta_2(\phi_2)$ is large enough so 
we 
can approximate these
functions by the forms shown in this figure.}  
\label{functions}
\end{figure}

The collision matrix  includes the following 
processes \cite{bdg} (Fig. \ref{functions}). 
 
{\bf \it(a) Capture,} $b_\alpha(\theta)$: 
rolling grains
are captured if the local angle of the sandpile $\theta(x,t)$ 
is smaller than the
generalized repose angle $\theta_\alpha(\phi_\beta)$.
The capture is proportional to  $R_\alpha$
\cite{cross-capture}. 
 
{\bf \it(b)
Amplification}, $a_\alpha(\theta)$: 
if the local angle $\theta(x,t)$ is larger than the generalized 
repose 
angle $\theta_\alpha(\phi_\beta)$, then some static grains of type $\alpha$
will be converted into rolling grains due a collision by  rolling
grains of type $\alpha$.
 The amplification rate
is proportional to the concentration $\phi_\alpha$ in the
sandpile, and to  $R_\alpha$. 

{\bf \it(c) Cross-amplification}, $x_\beta(\theta)$: when static grains
of 
type $\alpha$ are amplified by rolling grains of type $\beta$. This
cross-amplification occurs when the local angle $\theta(x,t)$ is larger
than $\theta_\beta(\phi_\beta)$.

Equation (\ref{bdg-r}) now reads 
\begin{equation}
\begin{array}{ll}
\displaystyle{
\frac{\partial R_\alpha}{\partial t}=-v_\alpha \frac{\partial
R_\alpha}{\partial x} }&
\displaystyle{
 + ~ \gamma_{\alpha\alpha}  ~ \Pi[\theta(x,t)-
\theta_\alpha(\phi_\beta)] ~ \phi_\alpha ~R_\alpha} \\
& 
\displaystyle{
- ~ \gamma_{\alpha\alpha} 
~ \Pi[\theta_\alpha(\phi_\beta)-\theta(x,t)]  ~R_\alpha}\\
& \displaystyle{
+ ~ \gamma_{\beta\alpha} ~ \Pi[\theta(x,t) -\theta_\beta(\phi_\beta)]  ~
\phi_\alpha  ~R_\beta}
\end{array}
\label{mcs-r}  
\end{equation}

\section{Steady-state solution.}
\label{ssregime}

We now calculate the steady-state solution of the equations of motion
for the two-species sandpile including the terms corresponding to
 cross-amplification $\mbox{M}_{\alpha\beta}$, $\alpha\neq\beta$, 
which were not considered in  \cite{makse2}.

We consider the geometry of a silo
of lateral size $L$ (Fig. \ref{coordinates}). We
assume that the difference 
$\psi = \theta_1(\phi_2) -
\theta_2(\phi_2) $
 is independent of the concentration $\phi_2$, then
$\psi = 
\theta_{11} - \theta_{21} = \theta_{12} - \theta_{22}$ 
(see Fig. \ref{angles}).
We set $v_1=v_2\equiv v$, and $\gamma_{11} = \gamma_{22} = \gamma$, and
we keep $\gamma_{12} \neq \gamma_{21}$.
We seek a solution  where the profiles of the sandpile
and of the
rolling grains do not change in time. Since stratification is an
oscillatory solution, stratification cannot be
observed for the steady-state solution. We set

\begin{mathletters}
\label{hr}
\begin{equation}
\frac{\partial h}{\partial t} = v ~ \frac{R^0}{L},
\label{h-steady}
\end{equation}
and 

\begin{equation}
\frac{\partial R_\alpha(x)}{\partial t}
= 0, 
\label{r-steady}
\end{equation}
\end{mathletters}
with boundary conditions
$R_\alpha (x=0) = R_\alpha^0,$ and $R_\alpha (x=L) =0$,
and $R^0\equiv R_1^0+R_2^0$.

We first calculate the profile of the total rolling species 
$R(x) = R_1(x) + R_2(x)$.
From (\ref{bdg-r}), (\ref{bdg-h}), 
and (\ref{hr})
we obtain
\begin{equation}
0 = \frac{\partial R(x)}{\partial t} = -v
\frac{\partial R(x)}{\partial x} -  v ~ \frac{R^0}{L}, 
\end{equation}
so that the profile of total rolling species decays linearly with $x$,

\begin{equation}
R(x)=  \frac{ R^0}{L} ( L - x).
\end{equation}

From (\ref{bdg-phi}), (\ref{mcs-r}), and (\ref{h-steady}) we obtain the concentrations

\begin{mathletters}
\begin{equation}
\label{c1}
{\displaystyle
\phi_1(x) = \frac{\gamma ~ \Pi[\theta_1(\phi_2)-\theta(x)] ~ R_1(x)}
{v R^0/L + \gamma \Pi[\theta-\theta_1] R_1(x) +  \gamma_{21}
\Pi[\theta-\theta_2] R_2(x)},
}
\end{equation}
\begin{equation}
\label{c2}
{\displaystyle
\phi_2(x) = \frac{\gamma ~ \Pi[\theta_2(\phi_2)-\theta(x)] ~ R_2(x)}
{v R^0/L + \gamma \Pi[\theta-\theta_2] R_2(x) +  \gamma_{12}
\Pi[\theta-\theta_1] R_1(x)}.
}
\end{equation}
\label{c}
\end{mathletters}

The equations for the rolling species are obtained from
(\ref{bdg-r}), (\ref{bdg-phi}), and
(\ref{hr}),
\begin{equation}
\label{eq2}
{\displaystyle 
v \frac{\partial R_\alpha(x)}{\partial x} =
 -(v R^0/L) ~ \phi_\alpha(x),
}
\end{equation}

Next, we divide the calculations in two regions: Region A, where
  $\theta_2(\phi_2)<\theta<\theta_1(\phi_2)$, and  Region B, where 
  $\theta<\theta_2(\phi_2)<\theta_1(\phi_2)$ (see Fig. \ref{angles}b).
The steady-state solution of Eqs.  (\ref{bdg-eq}) and (\ref{canonical2}) shows
 a strong  segregation pattern.  

{\bf Region A}. 
At the upper part of the pile
we find that  only
small grains are present (for 
$0\le x\le x_m$, with 
$x_m =R_1^0 L / R^0-v/(\gamma \psi)$, see below). 

If $\theta_2<\theta<\theta_1$,
 from (\ref{c})
we obtain \cite{solution1} 
\begin{eqnarray}
\phi_1(x)=1,& ~~~~~~~~ \phi_2(x)=0.
\label{phi-up}
\end{eqnarray}

Using (\ref{eq2}) and (\ref{phi-up})
we find the  profiles of the rolling species 
\begin{equation}
\label{r1}
R_1(x) = R^0_1 -\frac{R^0}{L} x, \>\>\>\> 
R_2(x) = R^0_2.
\end{equation}
The profile of the sandpile is obtained from (\ref{bdg-h}),
(\ref{mcs-r}),
 and (\ref{h-steady})

\begin{equation}
\label{roll1}
-\frac{v R^0}{L} = \gamma (\theta(x) - 
\theta_{11}) R_1(x) + \gamma_{21} (\theta(x) -
\theta_{21}) R_2(x), 
\end{equation}
so that
\begin{equation}
\label{theta1}
\theta(x)-\theta_{11} 
=\frac{ -v/\gamma - \psi \gamma_{21} L R^0_2/(\gamma R^0)}{L (R_1^0 +
R^0_2 \gamma_{21}/\gamma)/R^0 - x  }.
\end{equation}
This solution is valid when
$\theta(x)>\theta_1(\phi_2=1)=\theta_{21}$. 
Then it is valid
for  $x<x_m$, where $x_m$ is 
\begin{equation}
x_m = \frac{R_1^0}{R^0} L - \frac{v}{\gamma \psi}.
\end{equation}

{\bf Region B}. 
At the  lower part of the pile ($x_m \le x \le L$), we find that, after
a small region of the order of $v/(\gamma \psi)$,
mainly
large grains are present.

If $\theta<\theta_2<\theta_1$, from (\ref{bdg-h}), (\ref{mcs-r}), 
and (\ref{h-steady}) 
we obtain

\begin{equation}
-\frac{v R^0}{\gamma L} = [\theta(x) - \theta_{1}(\phi_2)] R_1(x) + 
 [\theta(x) -
\theta_{2}(\phi_2)] R_2(x),
\end{equation}
and therefore
\begin{equation}
\label{t2}
\theta(x) - \theta_2(\phi_2) = \frac{-v R^0 /(\gamma L) + \psi
R_1(x)}{R(x)}.
\end{equation}
Inserting  
  (\ref{t2}) in (\ref{c}) we obtain the concentrations as a function of
the rolling species

\begin{mathletters}
\begin{equation}
\label{c3}
{\displaystyle
\phi_1(x) = \frac{R_1(x)}{R(x)} \left ( 
1 + \frac{\gamma \psi L} {v R^0} R_2(x) \right ),
}
\end{equation}
\begin{equation}
\label{c4}
{\displaystyle
\phi_2(x) = \frac{R_2(x)}{R(x)} \left ( 
1 - \frac{\gamma \psi L} {v R^0} R_1(x) \right ).
}
\end{equation}
\label{cc}
\end{mathletters}
We obtain the equations for the rolling species
using Eqs. (\ref{eq2}) and  (\ref{cc}),

\begin{mathletters}
\begin{equation}
\label{q1}
{\displaystyle 
\frac{\partial R_1(x)}{\partial x} = - \frac{R_1(x)}{R(x)} \left(
\frac{ R^0}{L} + \frac{R_2(x)}{r} \right ),} 
\end{equation}
\begin{equation}
{\displaystyle 
\frac{\partial R_2(x)}{\partial x} = - \frac{R_2(x)}{R(x)} \left(
\frac{R^0}{L} - \frac{R_1(x)}{r} \right ),} 
\end{equation}
\end{mathletters}
\noindent
where  $r \equiv v/(\gamma \psi)$.
Setting $u=R_1/R$ we obtain from (\ref{q1})
\begin{equation}
\label{setting}
r u'=-(1-u) u,
\end{equation}
and the solution is
\begin{equation}
{\displaystyle
\frac{R_1(x)}{R(x)} = \frac{ 1} {1+C ~exp[(x-x_m)/r]}
},
\end{equation}
where  $C$ is an integration constant.
By considering the continuity at $x=x_m$, we obtain $C=R_2^0 L /(R^0 r)$
\cite{solution2}. 
The profile of the pile is obtained from (\ref{t2}) and (\ref{dependence})

\begin{equation}
{\displaystyle
\theta(x)-\theta_{22} 
=- m \phi_1(x) + \frac{ \psi R_1(x) L / R^0 - v/\gamma }{(L-x)}.
}
\label{theta2}
\end{equation}

We notice that the parameter $r=v/(\gamma\psi)$ is expected to be of 
the
order of the size of the grains, so for large system size 
$r\ll L$, and $C\gg 1$. Then
the 
steady-state solution
can be  simplified, and we arrive to the following simpler forms when
also considering that, as in the experiment,
an equal volume mixture is used (i.e., 
$R_1^0= R_2^0 = R^0/2$).

{\bf Region A}. Valid for
$0\le x\le x_m=L/2-v/(\gamma \psi)$.

\begin{mathletters}
\begin{eqnarray}
\phi_1(x)  =  1 &~~~~~~~~~~~&\phi_2(x) =  0 \\
R_1(x) = R^0(\frac{1}{2} -\frac{x}{L})& ~~~~~~~~~~~ &
R_2(x)= \frac{R^0}{2}\\ 
\theta(x)-\theta_{11} 
&=&\frac{- v/\gamma- \psi \gamma_{21} L / (2 \gamma)}{ (L/2)
(1+\gamma_{21}/\gamma) - x}.
\end{eqnarray}
\label{simple-up}
\end{mathletters}

{\bf Region B}. Valid $x_m \le x \le L$.

\begin{mathletters}
\begin{eqnarray}
\phi_1(x) & = & \exp[{-\frac{\gamma\psi}{v}(x-x_m)}] \\
R_1(x)& =& \frac{2 v}{\gamma \psi L} \phi_1(x)
 R (x) \\
\theta (x)-\theta_{22} &=& -m \phi_1(x)-\frac{v}{\gamma
(L-x)}.
\end{eqnarray}
\label{simple-down}
\end{mathletters}

\begin{figure}
\centerline{
\vbox{ \hbox{\epsfxsize=6.7cm
\epsfbox{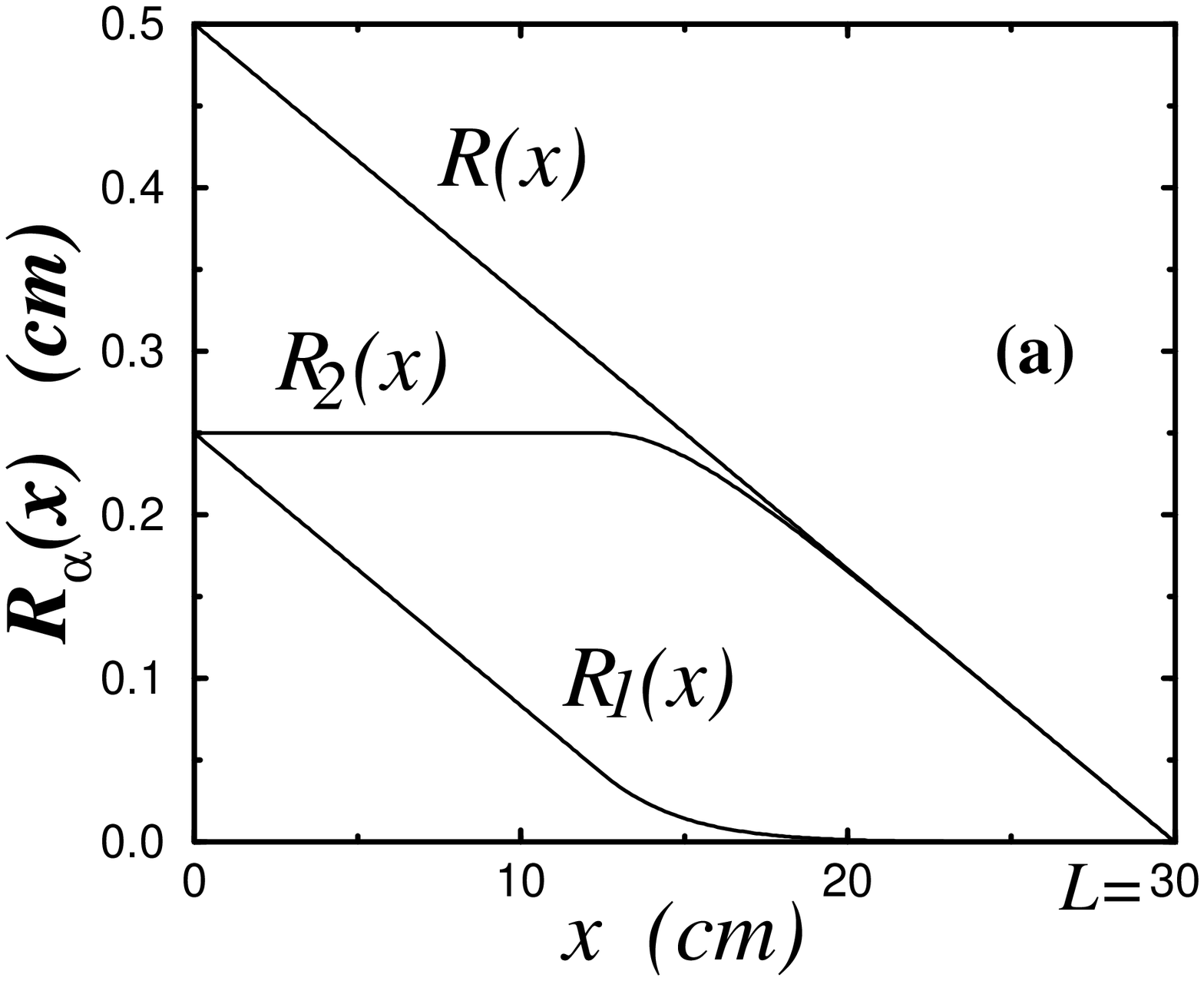}}
 \hbox{\epsfxsize=6.7cm 
\epsfbox{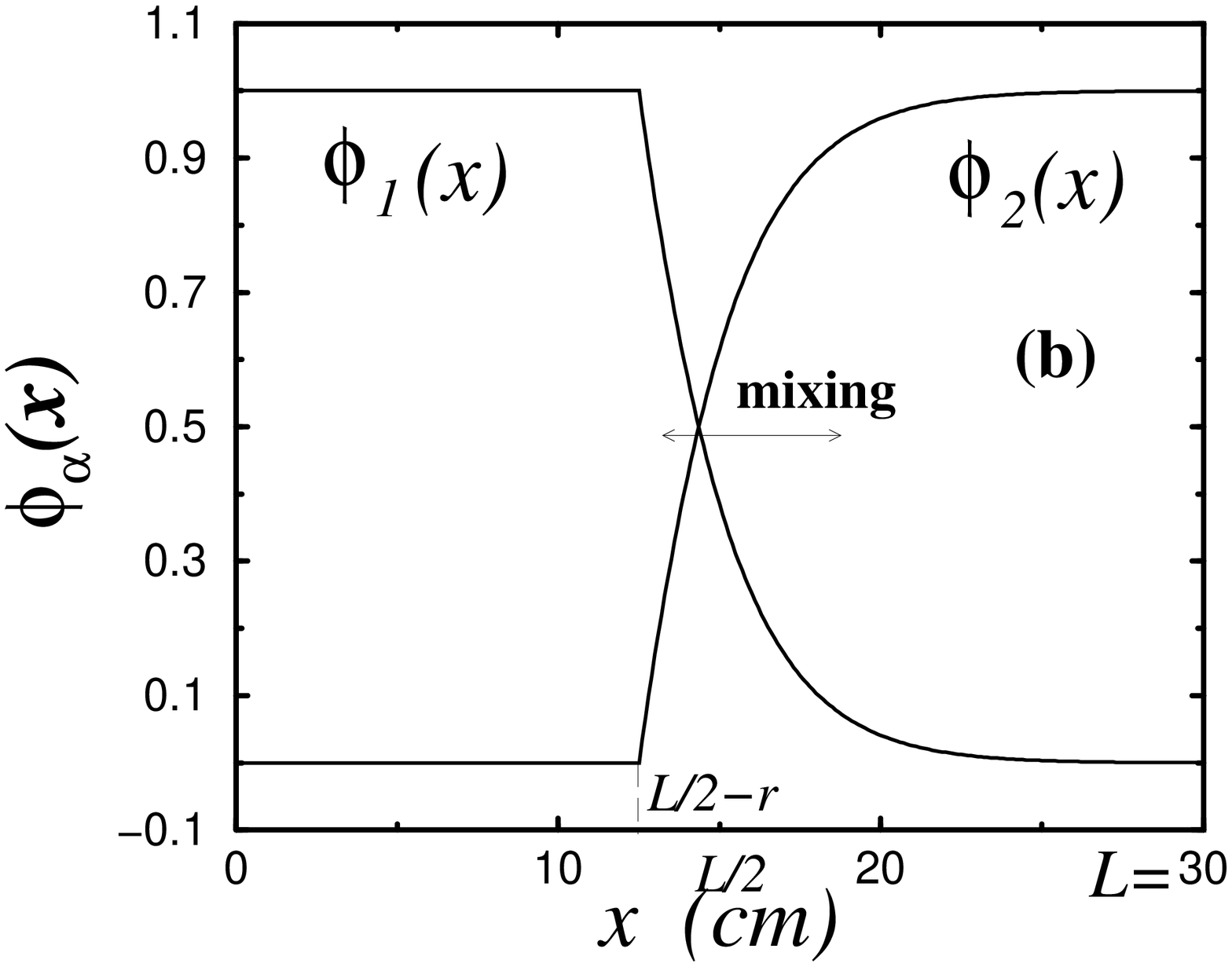} }
 \hbox{\epsfxsize=6.7cm 
\epsfbox{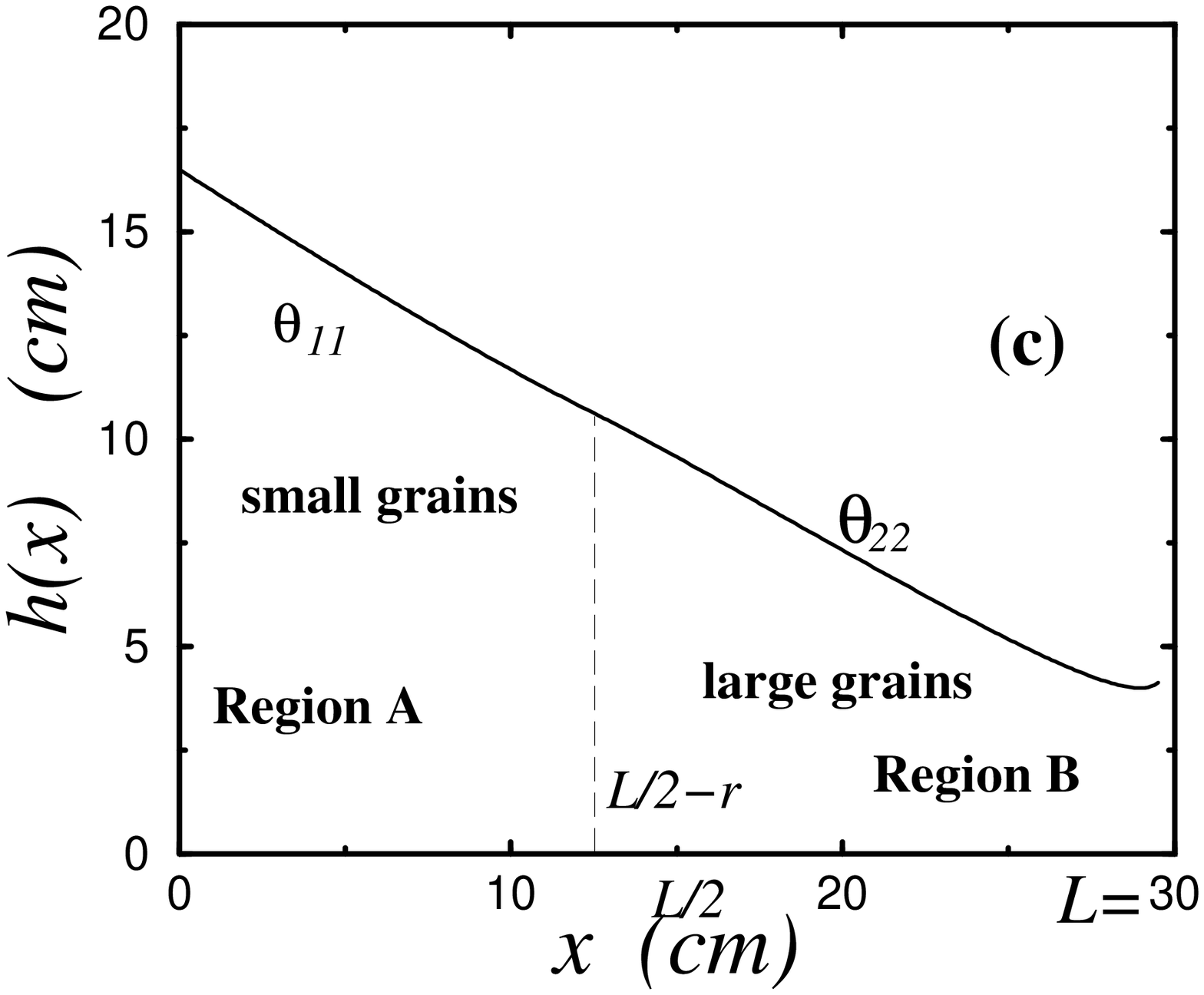} }
}
}
\narrowtext
\caption{Steady-state solution 
(\protect\ref{simple-up})--(\protect\ref{simple-down})  
for the
two-species granular flow in a silo
geometry. We use the following typical experimental
values for the phenomenological constants
appearing in the theory \protect\cite{difference}: $L = 30$ cm,
$\tan \theta_{11} = 0.6$, $\tan \theta_{22} = 0.5$, $\tan \psi =
0.2$, $\tan m = 0.1$,  $\tan \delta = -0.1$, $R^0 = 0.25$ cm, $v=10$
cm/sec,  $\gamma = 20$/sec, and $\gamma_{21} = 10$/sec.
{\bf a,} Profiles of the rolling species.
{\bf b,} Profile of the concentrations.
{\bf c,} Profile of the sandpile.}
\label{profiles}
\end{figure}

Figure \ref{profiles} shows the profiles
of the steady state solution
for typical experimental values. We use a cell of size 
$L=30$ cm. In \cite{makse4} the values of the different phenomenological
coefficients appearing in the theory were measured for a typical
equal-volume mixture
consisting of quasi-spherical glass beads of mean diameter $0.27$ mm,
and
cubic-shaped sugar grains of typical size $0.8$ mm.

The mean value of the velocity 
of the grains falling down the slope is of the order $v\simeq 10$ cm/sec.
The rate $\gamma$ was also estimated in \cite{makse4} to be of the order
of $\gamma \simeq 20$/sec. A typical value of the thickness of the 
layer of rolling grains is $R^0 \simeq 0.25$ cm \cite{rate}.
The difference $\tan\psi=\tan \theta_{11} - \tan \theta_{21}$ 
is of the order of $0.1-0.3$ \cite{difference}.
The concentrations (Fig. \ref{profiles}b) 
show the strong segregation of the mixture; the
mixing of the species 
is concentrated only in a small region, 
of the order of $v/(\gamma \tan\psi) \simeq 1.5-5$ cm, 
in the center of the pile.
The top part of the pile is made of small grains so that the angle 
is approximately equal to
$\theta_{11}$ (Fig. \ref{profiles}c).
Towards the center of the pile the angle decreases
and it is equal to $\theta_{21}$ 
at $x=x_m$ 
($\theta_{21}$ is the angle at which large grains start to be captured
on top of small grains). 
Then the angle increases gradually to $\theta_{22}$ towards the lower part
of the pile made of large grains.
The profile of $R(x)$  (Fig. \ref{profiles}a) behaves
linearly with $x$, which is a result of the 
conservation of number of grains Eq. (\ref{h-steady}). Same dependence is
observed in the case of a single-species sandpile \cite{pgg}.
The exponential behavior of the concentrations and rolling species
$R_\alpha(x)$ near
the center of the pile is expected, since we are solving equations of the
type
$\partial R_\alpha / \partial x \sim -(\gamma \psi/ v ) R_\alpha (x)$.
We notice that the segregation solution
(\ref{simple-up})-(\ref{simple-down}) is determined by the fact
that $\theta_1(\phi_2) > \theta_2(\phi_2)$, i.e. by the fact that 
the small grains are trapped first since they have a larger generalized
 angle of
repose
for a given concentration $\phi_2$. Then, this solution shows the
size-segregation of the mixture.

\section{Stability analysis}
\label{stability}

\begin{figure}
\centerline{
\vbox{  \hbox{\epsfxsize=8cm \epsfbox{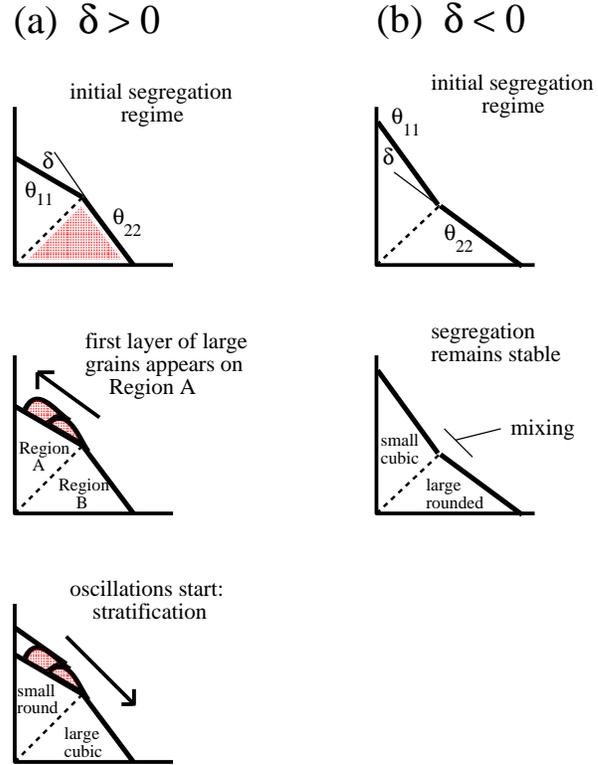}}
     }
       \vspace{1cm}    
}
\narrowtext
\caption{
 Schematic process
leading to {\bf a,} stratification ($\delta>0)$,  and {\bf b,} segregation
($\delta < 0$).}
\label{onset}
\end{figure}

In the experiments of \cite{makse1},
it was found that there is a 
transient regime of segregation before the layers
appear. 
The initial segregation regime turns into stratification only 
for mixtures with
$\delta >0$ (Eq. (\ref{delta})).
At the onset of
the instability leading to stratification, it is observed  
that a small amount of large grains is captured on top of the region of small
grains near the center of the pile (Region A, 
see Fig. \ref{onset}a).
This leads to the appearance of the first layer of 
larger grains and then to the oscillations characteristic of stratification.
 On the other hand, if $\theta_{11} >
\theta_{22}$ ($\delta < 0$) 
the segregation profile remains stable (Fig. \ref{onset}b).  
This picture was
also confirmed by the models proposed in \cite{makse2}.

The first sublayer of large grains corresponds to the appearance of the
first ``kink'' \cite{makse2,cms}.
The kink is an uphill wave at
which the  rolling grains are stopped. 
This first
``kink'' is formed only of large grains
(as was observed in \cite{makse1} and \cite{makse2}).
This kink appears to move uphill and creates an incipient layer of
large grains. Then 
the newly arriving grains roll down to the bottom of the pile 
where a new kink is developed.
At this point, the kink (formed by large and small grains) 
starts to move upwards and  all
grains are stopped at the kink, 
but the small-rounded grains are stopped first so
that the result is a pair of layers with the small-rounded grains being
found underneath the large-cubic grains.

Next,
we analyze the stability of the steady-state solution under
perturbations, by  assuming that the
steady-state solution obtained in Section \ref{ssregime}
is valid for the initial transient regime of
the evolution. We perturb the steady-state profiles by considering
that a small amount 
of large grains has been captured in the region A, near the center of the
pile
 (\(x \stackrel{<}{\sim} x_m\)), without
changing the angle of the pile.
We then analyze the short-time 
evolution of this perturbation to the steady-state profile.
The dynamical evolution of the additional  large rolling grains
are now  governed by the following equation:

\begin{equation}
\label{r2}
\frac{\partial R_2(x,t)}{\partial t} = - v \frac{\partial R_2}{\partial x}
+ \gamma (\theta - \theta_{22} ) R_2, 
\end{equation}
where the repose angle of the large rolling grains is $\theta_{22}$
since the surface is made only of large grains.

\noindent
We are interested 
in the behavior of the profiles near
the center of the pile, where $R_1$ is
very small, so we can focus only on the behavior of the large grains.
We look for the short-time behavior of $R_2$, so we can 
assume that the
angle of the pile remains unchanged from its initial value.
Then  
we replace $\theta(x,t)$ in (\ref{r2}) by $\theta(x)$ given by the
steady-state solution (\ref{theta1}), and
we  arrive to the following equation:
\begin{equation}
\label{r2b}
\displaystyle{
\frac{\partial R_2(x,t)}{\partial t} = - v \frac{\partial R_2}{\partial x}
- \left (\gamma \delta + \frac{v+v'}{(\ell-x)} \right ) R_2(x,t), }
\end{equation}
where $v'\equiv \psi \gamma_{21}  L R_2^0/R^0$, and $\ell\equiv L (R_1^0 +
R_2^0 \gamma_{21}  / \gamma )/ R^0$.
The solution of (\ref{r2b}) is
\begin{equation}
\label{solutionr2}
R_2(x,t) = ( \ell - x)^\omega ~ e^{-\gamma \delta t},
\end{equation}
with $\omega\equiv(v+v')/v$.

According to the exponential factor $e^{-\gamma \delta t}$
in (\ref{solutionr2}), 
$R_2(x,t)$  decreases as a function of time when $\delta > 0$.
This
implies that more large grains are captured, and the perturbation
of large grains evolves into the first sublayer of large grains and
then to stratification according to the experimental findings \cite{makse1}.
On the other hand, when  $\delta < 0$, $R_2(x,t)$  increases as a
function of time, so that the large grains of the initial perturbation
are amplified, the perturbation disappears, and the segregation 
profile remains stable.

\begin{figure}
\centerline{
\vbox{  \hbox{\epsfxsize=8cm \epsfbox{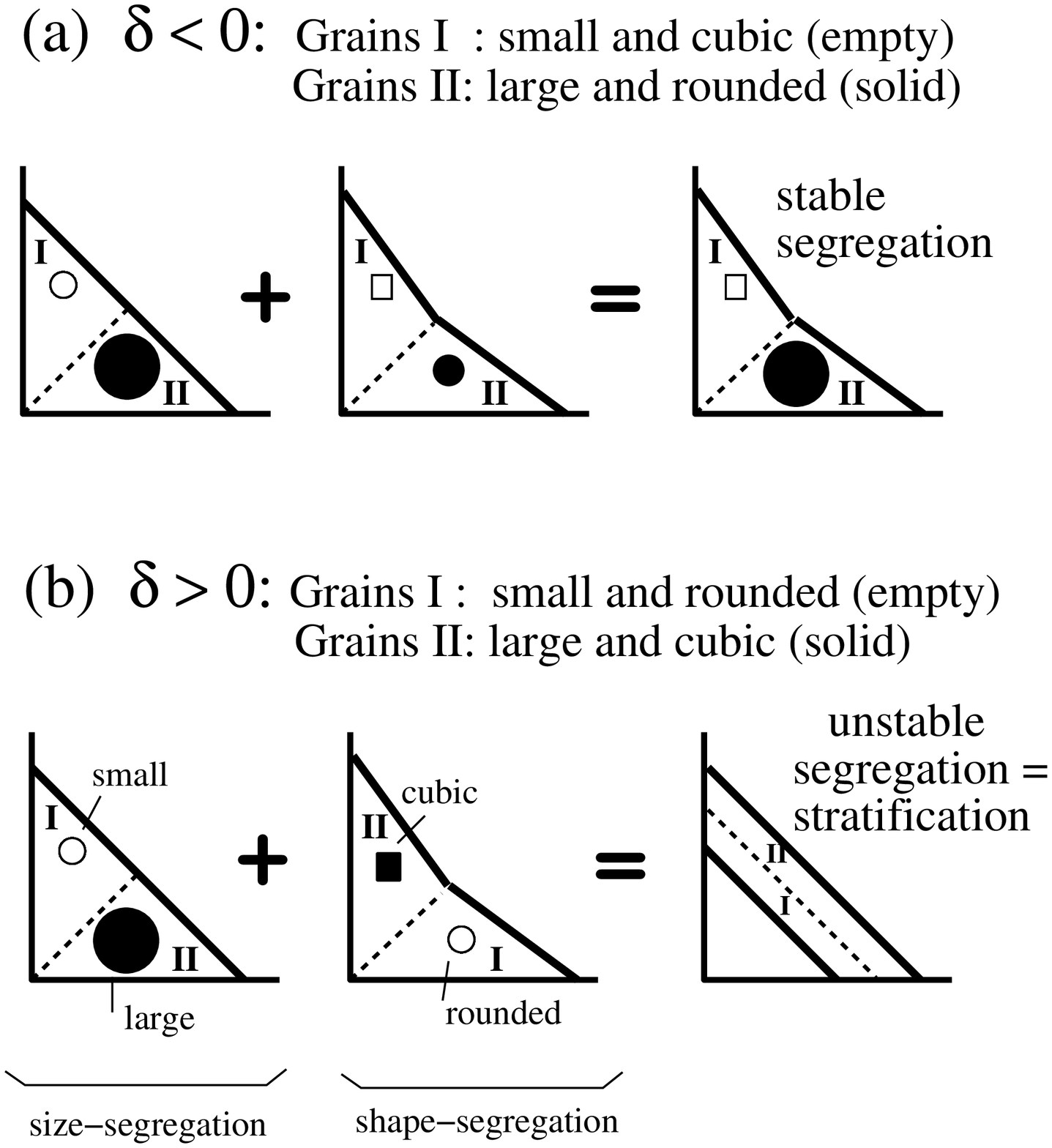}}
     }
       \vspace{1cm}    
}
\narrowtext
\caption{The instability mechanism for stratification
 is a result of 
two competing segregation effects.
Size-segregation (first panels in {\bf a}
and {\bf b}), and 
Shape-segregation (second panels in {\bf a} and {\bf b}). 
{\bf a,} Stable segregation when the mixture is composed 
of small-cubic grains
and large-rounded grains, and {\bf b,} unstable segregation leading to 
stratification when the
mixture is composed of small-rounded grains and large-cubic grains.} 
\label{why}
\end{figure}

\section{Stratification Instability}
\label{instability}

The steady-state solution 
calculated in Sec. \ref{ssregime}
shows the size-segregation of the mixture, while the stability
analysis of  Sec. \ref{stability}
shows that the grains also shape-segregate according to the value of
$\delta$.
The stable steady-state solution is
achieved when the mixture is composed of small-cubic
grains (Grains I in Fig. \ref{why}a) 
and large-rounded   grains  (Grains II  in Fig. \ref{why}a). 
In this case, the segregation of the mixture results
because size and shape-segregation 
act simultaneously to segregate the
large-rounded
grains at the bottom and the small-cubic  grains at the top.
On the other hand, 
when pouring a mixture of small-rounded grains
(Grains I in Fig. \ref{why}b) 
and large-cubic grains 
(grains II in Fig. \ref{why}b) 
an instability
develops  since the size-segregation and shape-segregation mechanisms 
tend to segregate the same grain in opposite regions of the cell.

\begin{figure}
\centerline{
\vbox{  \hbox{\epsfxsize=8cm \epsfbox{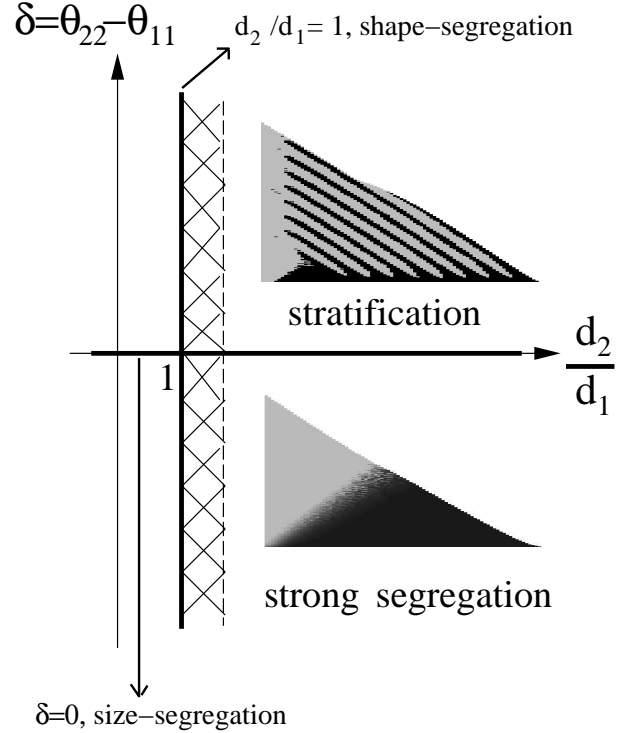}}
     }
       \vspace{1cm}    
}
\narrowtext
\caption{Phase diagram for two-dimensional 
flows of granular mixtures. 
$d_2$ is the size of the large grains, 
$d_1$ is the size of the small grains, $\theta_{22}$ is the repose angle
of the large grains, and  $\theta_{11}$ is the repose angle
of the small grains. We distinguish the  regions of stratification and strong
segregation (as shown by the steady-state solution presented here).
Weak segregation is expected in the shaded region 
when the size ratio is close to one. This case is discussed
in \protect\cite{bmdg}.} 
\label{phase}
\end{figure}

Qualitatively, the onset of the instability
can be seen as follows.
If a 
small amount of large grains
is captured near the center of the pile
where the angle of the pile is $\theta \simeq \theta_{11}$ (Point A in
Fig. \ref{angles}), then   the
repose angle for additional large rolling
grains is $\theta_{22}$.  Thus, if $\theta
\simeq \theta_{11} < \theta_{22}$ (Point B in Fig. \ref{angles}a), 
more large grains can be trapped
(since the angle of the surface is smaller than the repose angle),
leading to the first kink of large grains and then to
stratification.  On the other hand, the perturbation of large grains
 disappears when $\delta<0$. Since
 $\theta \simeq \theta_{11} >
\theta_{22}$ (Point C in Fig. \ref{angles}b), 
no more large grains are captured, the fluctuation
disappears, and the segregation profile remains stable.

In light of these results
we propose the phase diagram shown in Fig. \ref{phase}. The model
presented here
predicts a region of stratification $(\delta > 0)$, and a region of
strong segregation $(\delta<0)$ for size ratios not close to one. 
In Fig. \ref{phase} we show the results of numerical integration of the
equations of motion using the following parameters: 
stratification: $\delta = 0.25$, $\psi = 0.3$, $\gamma_{11}=\gamma_{22}
= 1$, $\gamma_{12}=\gamma_{21}=0.1$, $v_1=v_2=1$, $R_1^0=R_2^0=0.5$, and
segregation: the same parameters except for $\delta = -0.1$.
For other parameters such as $v_1 \neq v_2$, 
different $\gamma_{\alpha\beta}$, or 
$\theta_{12} - \theta_{11} \neq \theta_{22} - \theta_{21}$ 
we obtain similar results.
In addition, when the
cross-amplification rates $\gamma_{\alpha\beta}$ are of the order of
the amplification rates $\gamma_{\alpha\alpha}$ we also find 
oscillations in the center of the pile
which decay exponentially  when 
$\delta < 0$. 

\section{Discussion}
\label{discussion}

In summary, we 
study analytically  segregation and stratification in granular
mixtures focusing on the instability mechanism for stratification. 
We find the steady-state solution of the equations of motion
which shows strong
size-segregation with the small grains located at the top of the
pile,
and with the large grains located near the bottom of the pile.
In the center of the pile we find a region of mixing where the
concentration profiles behave exponentially with a
characteristic  region of mixing 
given by $v/(\gamma \psi)$. This region can be of the
order of $1.5-5$ cm and is observed experimentally.
We also find that 
 the steady-state
 solution is stable under a perturbation involving 
large grains trapped at the center-top of the pile, when 
$\delta<0$ (corresponding to large grains less faceted than
small grains). 
On the other hand, the steady-state solution is unstable under the same
perturbation
when $\delta>0$ (corresponding to large grains more faceted than
small grains).  The stratification instability is
related to the fact that, when $\delta>0$, there appears a competition
between
size-segregation
and shape-segregation:
the large-cubic 
grains tend to  size-segregate at the bottom of the pile, while at the 
same time, they tend to shape-segregate at the top of the pile.
Thus, the segregation solution becomes unstable, and the system
evolves spontaneously 
to  stratification.

ACKNOWLEDGEMENTS.
I would to thank T. Boutreux and P.-G. de Gennes for many illuminating 
discussions, and also the hospitality of the College de France where this
work was done.
I also thank
P. Cizeau,  H. J.
Herrmann, 
H. E. Stanley,
and S. Tomassone for stimulating discussions.
I also  acknowledge financial support from BP.

\end{multicols} 

\begin{references}
\bibitem{ratio1}
In addition, when the ratio between the size of the 
grains is close to one, only 
a weak segregation of the mixture
is found no matter the angle of
repose of the species \protect\cite{yan,makse4}.
\bibitem{ratio2}
We focus on the 
 case when the difference in  size of the species  is not too small 
(i.e. $d_2/d_1 > 1.4$, where $d_1$, $d_2$ are the typical size of
the small and large grains).
The case  $d_2/d_1 < 1.4$ give rise only to segregation
\protect\cite{yan,makse4} and is treated
in \protect\cite{bmdg}.
\bibitem{exchange}
We do not consider exchange processes \protect\cite{bdg}
in the definition of the collision
matrix.
\bibitem{cross-capture}
We notice that this term also includes ``cross-capture'', since a rolling
grain of type $\alpha$ can be captured on a surface of static grains of
type $\beta$ ($\phi_\beta = 1$).
\bibitem{solution1}
According to (\protect\ref{c}), another possible solution when
$\theta_2<\theta<\theta_1$ is obtained when the denominator of 
(\protect\ref{c1}) and (\protect\ref{c2}) is equal to $0$, and
$\phi_1\ne0$,  $\phi_2\ne0$. But this solution is not physical because, 
when $\theta>\theta_2$, large grains can only be
amplified but not captured,
so that $\phi_2 = 0$ is the only possible physical solution.
\bibitem{solution2}
Another possible solution is obtained by trying to imposed
the solution for Region B for all $x$. Then we do not consider Region A
as a possible solution and the constant $C$ is obtained from
$R_1(x=0) = R_1^0$. In this case $C=R_1^0/R_2^0$. However, this solution
is not physical since it would imply $\phi_1(x=0) > 1$ if $ R_1^0/R^0 >
r/L$.
\bibitem{rate}
This value was obtained
for a cell with a gap of $5$ mm between the vertical slabs, and pouring a
mixture at a rate of $10$ cm$^3$/sec.
\bibitem{difference}
We notice that for comparison with experimental results,
all the values of the angles appearing in the theory
should be replaced by the 
tangent of the angles (see Eq. \protect\ref{tangent}).
\end{references}
\end{document}